\documentclass[twocolumn]{jpsj2}

\def\runauthor{
H. {\sc Matsuura} et al,
}

\title{
Theory of Room Temperature Ferromagnet V(TCNE)$_x$ ($1.5 < x < 2$): \\
Role of Hidden Flat Bands
}          
\author{
Hiroyasu {\sc Matsuura}, Kazumasa {\sc Miyake} and Hidetoshi {\sc Fukuyama} $^{1}$
}
\inst
{
Division of Materials Physics, Department of Materials Engineering Science \\
Graduate School of Engineering Science, Osaka University, Toyonaka, 
Osaka 560-8531, Japan\\
$^{1}$Faculty of Science and Research Institute for Science and Technology (RIST)
Tokyo Univerisity of Science\\ 
1-3, Kagurazaka, Shinjuku-ku, Tokyo 162-8601, Japan
}

\recdate
{
\today
}

\abst
{Theoretical studies on the possible origin of room temperature ferromagnetism (ferromagnetic once crystallized) in the molecular transition metal complex, V(TCNE)$_x$ ($1.5<x<2$) have been carried out.
For this family, there have been no definite understanding of crystal structure so far because of sample quality, though the effective valence of V is known to be close to +2. 
Proposing a new crystal structure for the stoichiometric case of x=2, where the valence of each TCNE molecule is -1 and resistivity shows insulating behavior, exchange interaction among d-electrons on adjacent V atoms has been estimated based on the cluster with 3 vanadium atoms and one TCNE molecule. 
It turns out that Hund's coupling among d orbitals within the same V atoms and antiferromagnetic coupling between d oribitals and LUMO of TCNE (bridging V atoms) due to hybridization result in overall ferromagnetism (to be precise, ferrimagnetism).
This view based on localized electrons is supplemented by the band picture, which indicates the existence of a flat band expected to lead to ferromagnetism as well consistent with the localized view.
The off-stoichiometric cases ($x<2$), which still show ferromagnetism but semiconducting transport properties, have been analyzed as due to Anderson localization.
}

\kword{V(TCNE)$_x$, Flat band}
\begin{document}
\sloppy
\maketitle

\section{Introduction}
Recently, transition metal complexes with organic molecules have attracted much attention.
In particular, "ferromagnetism" was found to be realized in some compounds, 
such as T$_n$(C$_6$H$_6$)$_{n+1}$ (T=Sc, Ti, and V)~\cite{Nakajima} and V(TCNE)$_{x}$ (TCNE= tetracyanoethylene, $1.5 < x < 2$) ~\cite{Manriquez,Pokhodnya1,Miller}. 
The latter compound has been discussed intensively for the past two decades 
since its discovery in 1991~\cite{Manriquez},
 because of its high Curie temperature T$_c \sim 400$K, 
the highest of the transition-metal-organic-molecule complexes.
However, the mechanism of its ferromagnetism has not been clarified yet.
This is partly because the crystal structure of these materials has not been identified so far,
while many crystal structures of V(TCNE)$_x$ have been proposed.
On the basis of the band structure calculations, the stability of ferromagnetism was discussed~\cite{Tchougreeff2,Fusco}.
Nevertheless, the essence of the mechanism of ferromagnetism is not yet fully understood,
because these calculations treat many body effects only in one-electron approximation and then effects of strong correlations have not been fully assessed.

The purpose of this paper is twofold.
First, we propose a new structure of V(TCNE)$_x$ which satisfies all the experimental constraints.
Second, on the basis of this structure, 
1) the exchange interaction among electrons at adjacent V ions  is estimated by numerical diagonalization method which can take the electron correlation fully into account, and
2) the effect of the itinerancy of electrons is discussed by the tight-binding model,
which shows that a sharp peak of the density of states (DOS) at the Fermi level appears 
due to the existence of the flat band
when the couplings between two dimensional layers are absent.
The latter result can well explain the transport properties of the system 
exhibiting a resistivity of the variable-range-hopping behavior.     

\section{Electronic State of TCNE}
It is reported that the lowest unoccupied molecular orbital (LUMO) of TCNE strongly hybridizes with d orbitals of V~\cite{Kortright}.
The electronic state of LUMO has been discussed
 on the basis of the quantum chemical calculation~\cite{Erdin}.   
In this section, we show that LUMO is expressed qualitatively in the $\pi$ electron approximation.

\begin{figure}[h]
\begin{center}
\rotatebox{0}{\includegraphics[width=0.5\linewidth,angle=-90]{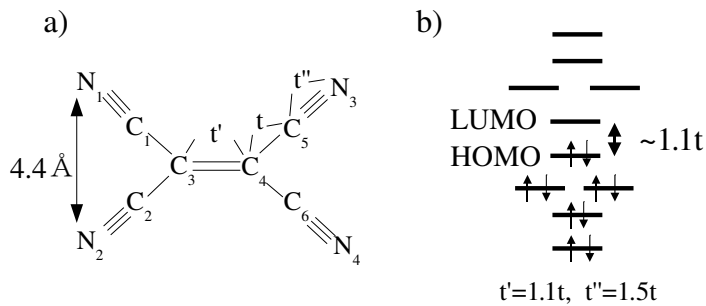}}
\caption{(a) Structure of TCNE. The distance between Ns is about 4.4$\AA$.
(b) Energy level scheme of the molecular orbitals of TCNE on the $\pi$ electron approximation. 
}
\label{TCNE}
\end{center}
\end{figure}

The structure of TCNE is shown in Fig. \ref{TCNE}(a)
 where C$_i$ and N$_i$ are i-th Carbon (C) and Nitrogen (N).  
Three transfer integrals $t$, $t^\prime$ and $t^{\prime\prime}$ among $\pi$-orbitals of C and N 
are shown in Fig. \ref{TCNE}(a). 
Fig. \ref{TCNE}(b) shows the energy level scheme of the molecular orbitals of TCNE
 in the tight-binding approximation for the resonable values of the transfer parameters $t^\prime=1.1t$ and $t^{\prime\prime}=1.5t$, because the transfer integrals are propotional to $d^{-2}$ where $d$ is the distance between C-C ($d\sim1.43\AA$), C=C ($d\sim1.37\AA$) and C$\equiv$N ($d\sim1.16\AA$).~\cite{Erdin}.
Then, the wave function of LUMO $\Phi_L$ is written as
$\Phi_{L} = A_1(-\varphi_{N_1}-\varphi_{N_2}+\varphi_{N_3}+\varphi_{N_4})+A_2(\varphi_{C_1}+\varphi_{C_2}-\varphi_{C_5}-\varphi_{C_6})+A_3(\varphi_{C_3}-\varphi_{C_4})$
where $\varphi_{N_i}$ and $\varphi_{C_i}$ are the wave functions of $\pi$-orbital of i-th N and C, respectively, and 
$A_1$ $\cong$ 0.355, $A_2$$\cong$ 0.129 and $A_3$ $\cong$ 0.463 are obtained with the same parameter set. 
LUMO is consistent with the wave function on the basis of the quantum chemical calculation~\cite{Erdin}.
LUMO has a large weight of $\pi$ orbitals of N.
Namely, it suggests the imporance of the hybridization between $\pi$ orbitals of N and d orbitals of V.  

It is also noted that the transfer integrals are generally propotional to $d^{-4}$~\cite{ryousi}.
At that time, these transfer integrals and the HOMO-LUMO gap are estimated as $t^{\prime} =1.2t$ and $t^{\prime\prime} =2.3t$, and of the order of $2t$.
The weight of C and N in LUMO is a little different from the case that the transfar integrals are propotinal to $d^{-2}$.
However, the energy level scheme of the molecular orbitals is qualitatively consistent with the case of $d^{-2}$. 

\section{Structure of V(TCNE)$_x$}
We propose a new structure of V(TCNE)$_x$ on the basis of experimental results. 
The coordination number of V in V(TCNE)$_x$ $\cdot$yS and V(TCNE)$_x$ has not been determined, 
because of the lack of x-ray diffraction lines due to the existence of defects of V~\cite{Miller}.
Nevertheless, the number of ligands around V has been studied in detail by EXAFS~\cite{Haskel}, 
showing that V has a slightly distorted octahedral environment constructed by six N atoms.
It was also proposed that these N atoms belong to different TCNEs 
from the measurement of the N-N distance of octahedra (2.8 $\AA$)  
which is shorter than the distance between two Ns in the TCNE molecule (4.4 $\AA$).

The nominal valence $n_V$ of V is estimated as $n_V=+2$ by XANES~\cite{Haskel}.
At that time, the $n_V/6$ electrons of V transfer to one TCNE, because V bonds with six TCNEs.
The charge neutrality requires that the valence $n_{TCNE}$ of TCNE should be $n_{TCNE}= -n_V/x$.
Then, the number $N_V$ of Vs bonding with TCNE is given as $N_Vn_V/6=n_V/x$, and corresponds to $N_V=3$ for $x=2$ and $N_V=4$ for $x=1.5$.

\begin{figure}[h]
\begin{center}
\rotatebox{0}{\includegraphics[width=0.4\linewidth,angle=-90]{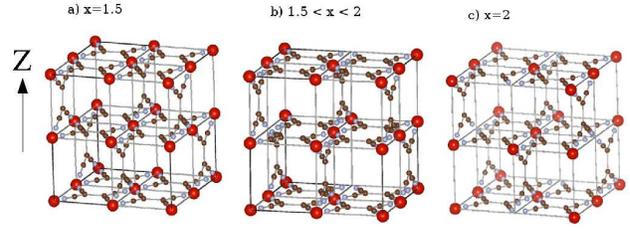}}
\caption{Structures of V(TCNE)$_x$.
Fig. 1(a), (b) and (c) are the structure of $x=1.5$, $1.5 < x < 2$ , and $x=2$, respectively.
Red, blown and blue circles indicate V, C and N, respectively. 
}
\label{V_TCNE_st2}
\end{center}
\end{figure}

Based on these experimental results, 
we propose the structure of V(TCNE)$_x$ as shown in Fig. \ref{V_TCNE_st2}:
Fig. \ref{V_TCNE_st2}(a) for V(TCNE)$_{1.5}$ and Fig. \ref{V_TCNE_st2}(c) for V(TCNE)$_2$. 
While the structure of Fig. \ref{V_TCNE_st2}(a) has already been proposed~\cite{Miller},
the structure of V(TCNE)$_{2}$ is new.
The structure of V(TCNE)$_x$, for $1.5 <x<2$, is expected to be an intermediate one 
between Fig. \ref{V_TCNE_st2}(a) and Fig. \ref{V_TCNE_st2}(c), which is obtained by extracting V ions from Fig. \ref{V_TCNE_st2}(a),
as shown in Fig. \ref{V_TCNE_st2}(b).
Note that Vs are randomly extracted from V(TCNE)$_{1.5}$, but 
all Vs have six-coordination. 

\section{Local Electronic State of V }
To construct an effective model of V(TCNE)$_x$, it is crucial to understand 
the local electronic state of V.
Since V is surrounded by six N atoms forming a slightly distorted octahedral structure,
we assume that the symmetry around V is D$_{4h}$, 
although the exact location of the apical N atoms has not been determined yet. 
The crystalline electric field (CEF) levels of the d orbitals of V depends on the location of apical N atoms.

From the symmetry requirement, the hybridizations are allowed 
only between $d_{yz}$ and $d_{zx}$ orbitals of V and $\pi$ orbitals of N.
Here, we have defined the x and y axes so that N atoms of the V-TCNE plane will lie on the x and y axes, and V atoms at the origin.
We also have defined the z axis so that the apical N atom will lie on the z axis. 
On the other hand, the $d_{xy}$ orbital of V can be regarded as an almost localized one.
We also find that $e_g$ orbitals of V do not hybridize with $\pi$ orbitals, but hybridize with $\sigma$ orbitals of N.
The $\sigma$ bonding formed between these orbitals makes a strong covalent bonding.~\cite{Kortright}.
Namely, these bonding orbitals are important for stabilizing the structure of V(TCNE)$_x$.

\section{Cluster Model - Estimate of Exchange Intercation}
\begin{figure}[h]
\begin{center}
\rotatebox{0}{\includegraphics[width=0.4\linewidth,angle=-90]{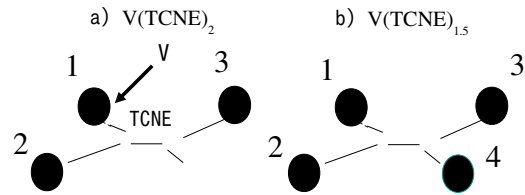}}
\caption{(a) Cluster of three Vs and TCNE. (b) Cluster of four Vs and TCNE.
}
\label{Fig3}
\end{center}
\end{figure}

Here, we construct an effective cluster model for $x=2$ focusing 
on three Vs and TCNE as shown in Fig. \ref{Fig3}(a).
We also discuss the case of x=1.5 for four Vs and TCNE as shown in Fig. \ref{Fig3}(b).
We will estimate the exchange interaction among electrons at two adjacent V atoms.
The Hamiltonian for the case of Fig. \ref{Fig3}(a) consists of three parts, 
the hybridization part $H_{\rm{0}}$, the $d$-site part $H_{\rm{d}}$, and the LUMO part $H_{\rm{LUMO}}$ as 
\begin{equation}
H = H_{\rm{0}} + H_{\rm{d}} + H_{\rm{LUMO}}.
\end{equation}
Explicit forms of these terms are written as follows:
\begin{eqnarray}
H_{\rm{0}} &=& -t\sum_{\sigma}( d_{(1,2)\sigma}^{\dag} -d_{(2,1)\sigma}^{\dag} - d_{(3,1)\sigma}^{\dag} )\varphi_{\sigma} + h.c. \nonumber \\
                && + \Delta_{d}\sum_{i=1}^{3}\sum_{\sigma}n_{(i,3)\sigma}^d +\Delta_{L}\sum_{\sigma} n_{\sigma}^L  \label{Hamilt}\\
H_{\rm{d}}  &=& 
       + U_d \sum_{i=1}^{3} \sum_{j=1}^{3} n_{(i,j)\uparrow}^dn_{(i,j)\downarrow}^d \nonumber \\  
       && +\frac{1}{2}(U_d^\prime -J_d)\sum_{i=1,\sigma}^{3} \sum_{j,k=1 (j\neq k)}^{3}n_{(i,j)\sigma}^dn_{(i,k)\sigma}^d \nonumber \\
  & & +\frac{1}{2}U_d^\prime\sum_{i,\sigma\neq \sigma^\prime}^{3} \sum_{j,k=1 (j\neq k)}^{3}n_{(i,j)\sigma}^dn_{(i,k)\sigma^\prime}^d \nonumber \\
 &&+\frac{1}{2}J_d \sum_{i}^3\sum_{j,k=1 (j \neq k))}^{3}\biggr( d_{(i,j)\uparrow }^{\dag}d_{(i,k)\uparrow }
d_{(i,k)\downarrow }^{\dag}d_{(i,j)\downarrow } + {\rm h.c.}\biggr)  \label{interac1} \\
H_{\rm{LUMO}} &=&  U_{L}n_{\uparrow}^L n_{\downarrow}^L.   \label{interac2}
\end{eqnarray}
Here, d$_{(i,j) \sigma}$ is an annihilation operator of an electron with spin $\sigma$ on the $d$ orbital $j$ at site i 
;d$_{(i,1) \sigma}$, d$_{(i,2) \sigma}$ and d$_{(i,3) \sigma}$ represent $d_{zx}$, $d_{yz}$ and $d_{xy}$, respectively ,while $\varphi_{\sigma}$ is that on LUMO;
$t$ is the transfer integral between $d$ orbitals and LUMO; 
$\Delta_ d$, U$_d$, U$_{d}^\prime$ and J$_d$ are one-body level of $d_{xy}$ orbital,  intra-, inter-, and exchange Coulomb interaction parameters;
$\Delta_{L}$ and U$_{L}$ are one-body level of LUMO and Coulomb interaction parameter of LUMO.

\begin{figure}[h]
\begin{center}
\rotatebox{0}{\includegraphics[width=0.3\linewidth,angle=-90]{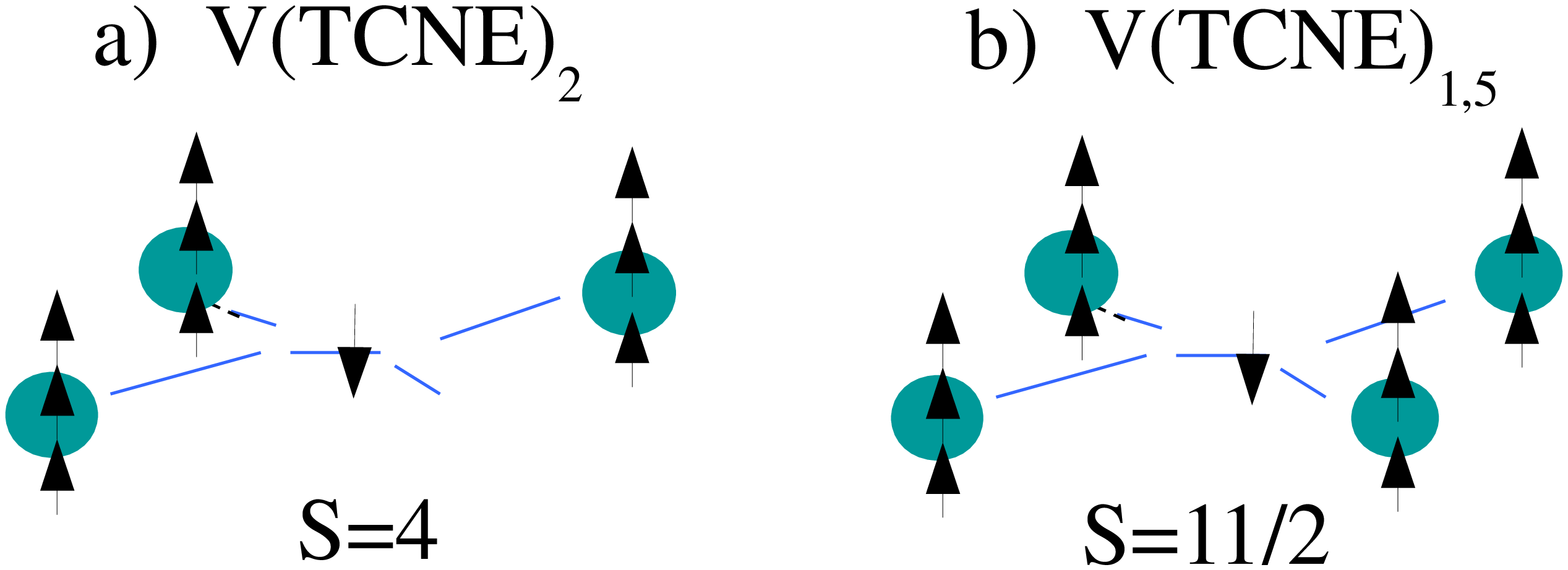}}
\caption{Snapshot of the spin states of (a) V(TCNE)$_2$, and (b) V(TCNE)$_{1.5}$ on the valence bond picture.
}
\label{V_TCNE}
\end{center}
\end{figure}

In the case of V(TCNE)$_2$, each V has three electrons on d orbitals and TCNE has one electron on LUMO.
Thus, the total number of electrons is 10 ($=3\times3 + 1$) in the truncated model. 
We have studied the ground state of this model of the cluster of V(TCNE)$_2$ on the basis of the numerical diagonalization method.
It turns out that the ground state of the cluster of V(TCNE)$_2$ has S=4  state as shown in Fig. \ref{V_TCNE}(a) for the parameter set 
$U_d/t=4.0$, $U_d^\prime/t=3.0$, $J_d/t=0.5$, $U_L/t=2$, $\Delta_d/t =0.0$ and $\Delta_L/t=2.0$.
Note that the ground state of $S=4$ prevails in a wide region of the parameter space
 of $\Delta_L/t$ and $\Delta_d/t$ although it is not shown here.
For the parameter set $\Delta_d/t=0$ and $\Delta_L/t =2.0$, 
the effective exchange coupling $J$ between d orbitals is estimated as 
$2J/t = E_{S=3}/t -E_{S=4}/t$ $\simeq$ 0.08,
where $E_{S=3}$ and $E_{S=4}$ are the energies of  
the first excited state and the ground state.

The nature of the spin state of $d_{(1,yz)}$ orbitals of V and LUMO of TCNE can be seen from 
the spin$-$spin correlations $\langle {s}_{yz}^z {s}_{\varphi}^z \rangle$ 
between $d_{(1,yz)}$ and $\varphi$, and 
 $\langle {s}_{yz}^z {s}_{zx}^z \rangle$ between $d_{(1,yz)}$ and $d_{(1,zx)}$ or $d_{(1,xy)}$. 
They are estimated as $\langle {s}_{yz}^z {s}_{\varphi}^z \rangle$ $\simeq$ $-0.067t$ and 
$\langle {s}_{yz}^z {s}_{zx}^z \rangle$ $\simeq$ $0.18t$ for the same parameter set.
These results imply that the antiferromagnetic correlation works between $d_{(1,yz)}$ 
 and $\varphi$, and the ferromagnetic correlation works between $d_{(1,yz)}$ 
 and $d_{(1,zx)}$ or $d_{(1,xy)}$.

The origin of the ferromagnetic exchange between $d$ orbitals at adjacent sites 
is understood on the basis of the valence bond picture as follows:
due to the kinetic energy gain,  
spin correlation between electrons in $d_{(1,yz)}$ and LUMO becomes antiferromagnetic 
as in the super-exchange correlation.
Similarly, spin correlation between electrons in $d_{(2,zx)}$ (or $d_{(3,zx)}$) and LUMO becomes antiferromagnetic, 
causing a ferromagnetic correlation between $d_{(1,yz)}$ and $d_{(2,zx)}$ (or $d_{(3,zx)}$) orbitals.
Therefore, electrons hopping among $d_{yz}$ (or $d_{zx}$) orbitals and LUMO 
takes the spin state, $S=1$ in the truncated system and $S=0$ in the full lattice systems of $d_{yz}$ and $d_{zx}$ orbitals and LUMO of V(TCNE)$_2$.
On the other hand, spins of $d_{(1,yz)}$, $d_{(1,zx)}$ and $d_{(1,xy)}$ 
become parallel due to the Hund's rule coupling as shown in Fig. \ref{V_TCNE}(a).
Then, considering  t $\sim$ 1eV, the Curie temperature $T_c$, which will be of the order of effective exchange coupling, $J$, is estimated to be of the order of 1000(K), which is consistent with experimental values of $T_c$.

This mechanism of the ferromagnetic correlation on the valence bond picture, 
in which electrons in LUMOs have opposite spin polarization of $d_{yz}$ and $d_{zx}$ orbitals
 by the $\pi$-d hybridization,   
 corresponds to the mechanism which Kanamori and Terakura 
discussed on the band picture.~\cite{Kanamori1}
In other words, the magnetically ordered state can be regarded as a kind of ferrimagnetism
 as shown in Fig. \ref{V_TCNE}(a).
Such a ferrimagnetic nature has been pointed out by L. Valade and her coworkers on the basis of measurement of saturated magnetic moment~\cite{Valade}.

\begin{figure}[h]
\begin{center}
\rotatebox{0}{\includegraphics[width=0.5\linewidth,angle=-90]{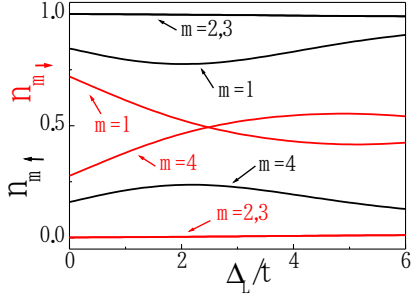}}
\caption{
Number of electrons of the up spin (black line) or down spin (red line) in MOs of the cluster of V(TCNE)$_{2}$ with the set of the parameters, $U_d/t=4.0, U_d^\prime/t=3.0, J_d/t=0.5, \Delta_d/t=0.0$, and $U_L/t=2.0$.
Index m is the number of the molecular orbitals.}
\label{Number1}
\end{center}
\end{figure}

The origin of this ferromagentism may be understood based on the molecular orbital picture as follows.
The molecular orbitals $\Phi_1 \sim \Phi_4$ of the cluster of V(TCNE)$_{2}$, which are the eigenstates of the Hamiltonan $H_0$ eq. (\ref{Hamilt}), are given, in order of increasing eigen values, by
\begin{eqnarray}
\Phi_1 &=& \alpha(d_{(1,yz)}  + d_{(2,zx)} + d_{(3,zx)}) +\beta \varphi    \\
\Phi_2 &=& \frac{1}{\sqrt{2}}( d_{(2,zx)} - d_{(3,zx)})                 \\
\Phi_3 &=& \frac{1}{\sqrt{2}}( d_{(1,yz)} - d_{(3,zx)})                  \\
\Phi_4 &=& \alpha^\prime(d_{(1,yz)}  + d_{(2,zx)} +  d_{(3,zx)}) + \beta^\prime \varphi,  
\end{eqnarray}
where $\alpha$, $\beta$, $\alpha^\prime$ and $\beta^\prime$ are the normalization factors, and  $\alpha \neq \alpha^\prime$ and $\beta\neq \beta^\prime$. 
The molecular orbitals $\Phi_1$ and $\Phi_4$ are the bonding and antibonding molecular orbitals.  
The orbitals $\Phi_2$ and $\Phi_3$ are 
non-bonding molecular orbitals (NBMOs) consisting of only d components.
Without interactions, $\Phi_1$ and $\Phi_2$ or $\Phi_3$ ($\Phi_2$ and $\Phi_3$ are degenerate) are doubly occupied.
In the presence of interactions, eq. (\ref{interac1}) and (\ref{interac2}), the occupation number of $\Phi_1$ $\sim$ $\Phi_4$ orbitals is altered so as to minimize the total energy.
Fig. \ref{Number1} shows the result of the number of electrons in the molecular orbitals $\Phi_1 \sim \Phi_4$ as a function of $\Delta_L/t$ 
in the Hilbert space of $S_z=4$ on the basis of the numerical diagonalization method 
with the same parameter set as $U_d/t=4.0$, $U_d^\prime/t=3.0$, $J_d/t=0.5$, $U_L/t=2$, $\Delta_d/t =0.0$ and $\Delta_L/t=2.0$.
Here $n_{m\uparrow}$ ($n_{m\downarrow}$) indicates the number of the up spin (down spin) electrons of the m-th molecular orbital.
The number of electrons in NBMOs ($m=2$ and $3$) is almost equal to one.
Furthermore, the electrons in NBMOs are in a nearly full spin polarized state, i.e. "ferromagnetic state".
On the other hand, the total numbers of electrons in $\Phi_1$ and $\Phi_4$ are nearly two and S$_z$ $\sim$ 0.
 The origin of the ferromagnetic correlation is also understood by the same way of the valence bond picture.

In the case of V(TCNE)$_{1.5}$, in which the total number of electrons is 13 ($=3\times4 +1$), 
the ground state has $S=11/2$ state
 in the truncated model such as V(TCNE)$_2$.
These results for the cluster of V(TCNE)$_{1.5}$ show that the origin of ferromagnetic correlations of V(TCNE)$_{1.5}$
 is the same as in V(TCNE)$_2$.

Precisely speaking, the mechanism of the ferromagentism discussed here
 can be applied only to the insulating state 
where the electrons are localized on each orbital.
However, V(TCNE)$_x$ (1.5 $<$ x $<$ 2) exhibits a temperature dependence on 
the conductivity $\sigma \propto T^{-1/2}\exp[-(T_0/T)^{1/4}]$ with $T_0 = 2.0 \times 10^{9}K$ 
described by the variable range hopping (VRH) mechanism~\cite{Prigodin}.
This implies that these systems are in the Anderson insulator due probably to the strong randomness with 
the localization length $\xi_{loc}$, far longer than the unit cell of V(TCNE)$_x$.
At that time, we have also to take into account the effect of itinerancy on the basis of band picture.

\section{Band Model - Characteristics of Crystal Structure} 
The feature of the structures of V(TCNE)$_x$ as shown in Fig.2 is 
that they have planes of V and TCNE with rare numbers of pillars of TCNE molecules connecting these planes.
The electronic state of this 2-d structure of V(TCNE)$_2$ will be discussed on the basis of the tight binding (TB) approximation, by taking into account only $d_{yz}$, $d_{zx}$ orbitals and LUMO of TCNE 
as in the case of the cluster model.
\begin{figure}[h]
\begin{center}
\rotatebox{0}{\includegraphics[width=0.5\linewidth,angle=-90]{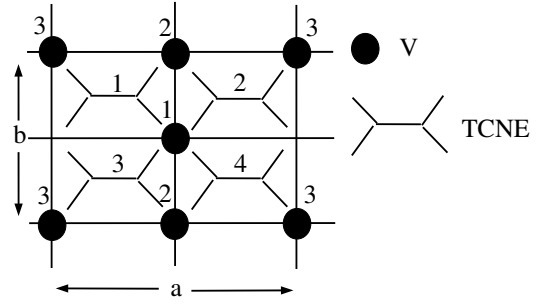}}
\caption{Two dimensional plane of V(TCNE)$_2$. 
}
\label{Fig:6}
\end{center}
\end{figure}  
The model Hamiltonian H$_{2d}$ of the 2-d plane of V(TCNE)$_2$ is written as
\begin{eqnarray}
&&H_{2d} = \sum_{{\bf k}}\biggr(  
\epsilon_{1{\bf k}}d_{(1,zx){\bf k}}^{\dag}\varphi_{2{\bf k}}  +
{\epsilon_{1{\bf k}}}^*d_{(1,zx){\bf k}}^{\dag}\varphi_{3{\bf k}}  \nonumber \\
&&+\epsilon_{2{\bf k}}d_{(1,yz){\bf k}}^{\dag}\varphi_{1{\bf k}} 
+{\epsilon_{2{\bf k}}}^*d_{(1,yz){\bf k}}^{\dag}\varphi_{4{\bf k}}  \nonumber \\
&&+{\epsilon_{1{\bf k}}}^*d_{(2,zx){\bf k}}^{\dag}\varphi_{1{\bf k}}  
+{\epsilon_{1{\bf k}}}d_{(2,zx){\bf k}}^{\dag}\varphi_{4{\bf k}}  \nonumber \\
&&+{\epsilon_{2{\bf k}}}^*d_{(2,yz){\bf k}}^{\dag}\varphi_{2{\bf k}}  
+{\epsilon_{2{\bf k}}}d_{(2,yz){\bf k}}^{\dag}\varphi_{3{\bf k}}   \nonumber \\
&&+{\epsilon_{1{\bf k}}}^*d_{(3,zx){\bf k}}^{\dag}\varphi_{2{\bf k}}  
+{\epsilon_{1{\bf k}}}d_{(3,zx){\bf k}}^{\dag}\varphi_{3{\bf k}}   \nonumber \\
&&+{\epsilon_{2{\bf k}}}^*d_{(3,yz){\bf k}}^{\dag}\varphi_{1{\bf k}}  
+{\epsilon_{2{\bf k}}}d_{(3,yz){\bf k}}^{\dag}\varphi_{4{\bf k}}   +h.c. \biggr) \nonumber \\
&& +\Delta_{L}\sum_{i=1}^4 \varphi_{i{\bf k}}^{\dag}\varphi_{i{\bf k}},
\end{eqnarray}
where $\Delta_L$ is the energy level of LUMO relative to the d-orbitals, and  * denotes taking complex conjugation, with $\epsilon_{i{\bf k}}(i=1,2)$ given as
\begin{eqnarray}
\epsilon_{1{\bf k}} &=&  t\exp(-ik_x a/4 -ik_y b/4), \\
\epsilon_{2{\bf k}} &=& -t\exp( ik_x a/4 -ik_y b/4), 
\end{eqnarray}
where $a$ and $b$ are lattice constants as shown in Fig. \ref{Fig:6}.
The numbering of the operators of V and TCNE is shown in Fig. \ref{Fig:6}.
Here, the terms of $d_{xy}$ orbital are ignored, because these localized orbitals are decoupled from $H_{2d}$.

The matrix of each ${\bf k}$ of this Hamiltonian (3) is given as two matrices of $5 \times 5$ as follows:
\begin{eqnarray}
 \bordermatrix{
                       &      d_{(1,zx){\bf k}} &d_{(2,yz){\bf k}} &d_{(3,zx){\bf k}} &\varphi_{2{\bf k}} &\varphi_{3{\bf k}}   \cr     
d_{(1,zx){\bf k}}^\dag  & 0                     &   0             &  0              &  \epsilon_{1{\bf k}}       &\epsilon_{1{\bf k}}^{*}  \cr
d_{(2,yz){\bf k}}^\dag  & 0                     &   0             &  0              & \epsilon_{2{\bf k}}^{*} &\epsilon_{2{\bf k}}   \cr
d_{(3,zx){\bf k}}^\dag  & 0                     &   0             &  0              & \epsilon_{1{\bf k}}^{*} &\epsilon_{1{\bf k}}    \cr
\varphi_{2{\bf k}}^\dag&\epsilon_{1{\bf k}}^{*} & \epsilon_{2{\bf k}} & \epsilon_{1{\bf k}}   &  \Delta_L                     &  0        \cr
\varphi_{3{\bf k}}^\dag&\epsilon_{1{\bf k}} & \epsilon_{2{\bf k}}^{*} & \epsilon_{1{\bf k}}^{*}   &   0         & \Delta_L   \cr
}, \label{eq1}
\end{eqnarray} 
\begin{eqnarray}
 \bordermatrix{
                       &      d_{(1,yz){\bf k}} &d_{(2,zx){\bf k}} &d_{(3,yz){\bf k}} &\varphi_{1{\bf k}} &\varphi_{4{\bf k}}  \cr
d_{(1,yz){\bf k}}^\dag  & 0                     &   0             &  0              &  \epsilon_{2{\bf k}}       &\epsilon_{2{\bf k}}^{*}  \cr
d_{(2,zx){\bf k}}^\dag  & 0                     &   0             &  0              &  \epsilon_{1{\bf k}}^{*} &\epsilon_{1{\bf k}}   \cr
d_{(3,yz){\bf k}}^\dag  & 0                     &   0             &  0              &  \epsilon_{2{\bf k}}^{*} &\epsilon_{2{\bf k}}    \cr
\varphi_{1{\bf k}}^\dag& \epsilon_{2{\bf k}}^{*}  & \epsilon_{1{\bf k}}  &  \epsilon_{2{\bf k}}              &  \Delta_L                     &  0        \cr
\varphi_{4{\bf k}}^\dag& \epsilon_{2{\bf k}}    &  \epsilon_{1{\bf k}}^{*}  &  \epsilon_{2{\bf k}}^{*}    &  0                               & \Delta_L   \cr
}.  \label{eq2}
\end{eqnarray} 
One of these eigenvalues of eq.(\ref{eq1}) or eq.(\ref{eq2}) is zero  
independent of wave vector ${\bf k}$ and $\Delta_L$.
Namely, there exist the flat bands, since the structure of this matrix is the same as the superhoneycomb lattice~\cite{Shima}. 

\begin{figure}[h] 
\begin{center}
\rotatebox{0}{\includegraphics[width=0.5\linewidth,angle=-90]{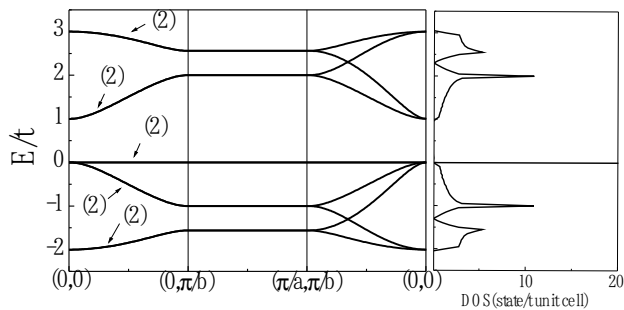}}
\caption{Band structure of V(TCNE)$_2$.
The number in () indicates the degeneracy of each orbitals.
}
\label{Fig:7}
\end{center}
\end{figure}

Fig. \ref{Fig:7} shows the result of the band structure along the symmetry axis in 
$\bf{k}$ space and the density of the states (DOS) for a choice of parameter of $\Delta_L/t=1.0$.
At $E/t =0$, two degenerate flat bands are found leading to the delta function in DOS.
The wave function of the flat bands are constructed only from d components.
Namely, of six orbitals of d character, two linear combinations of d-orbitals are decoupled from 
$\pi$-d hybridization bands such as the NBMOs in the cluster model, which is the origin of the flat band. 

In the case of $\Delta_d/t >0 $, where $\Delta_d$ is the energy level of d$_{xy}$ orbital, 
the Fermi level is located on the flat bands. 
A possible ferromagnetism associated with a flat band has been extensively 
discussed previously.~\cite{Tasaki}

In the case of 2-d version of V(TCNE)$_{1.5}$, 
a flat band is also found on the basis of TB approximation. 
In this case, the occupation of the flat band is incomplete.
Therefore, the same mechanism as V(TCNE)$_2$ is expected to stabilize the ferromagnetism.

In the actual system, however, the flat band is smeared to some extent
because of three dimensionality and impurity effects.
Nevertheless, the large DOS is still expected at the Fermi level.
Indeed, from the photoemission spectroscopy experiment, the exsistence of the large DOS of $d$ orbital component has been observed near the fermi level.~\cite{Tengstedt} 
In the theoretical description,  the DOS for the structure of Fig. 2(c) has the very same aspects as shown in Fig. 8.
Here, the effect of pillars of TCNE is taken into account properly.
Note that there remains huge (technically divergent) DOS at the Fermi level, which is consistent with the experimental result, 
even for the reasonable value of the transfer parameter $t^\prime_{pillar} = 0.5t$
between d$_{yz}$ (or d$_{zx}$) orbital and LUMO of TCNE of pillar. 
The origin of the peak structure of DOS around the Fermi level 
can be traced back to the flat band in the two dimensional sector shown in Fig. 6.   
Therefore,  
the ferromagnetism in V(TCNE)$_x$ ($1.5 < x < 2$) can be explained fairly well by the mechanism which Kanamori proposed for the ferromagnetism of Ni metal.~\cite{Kanamori2}

\begin{figure}[h]
\begin{center}
\rotatebox{0}{\includegraphics[width=1\linewidth,angle=0]{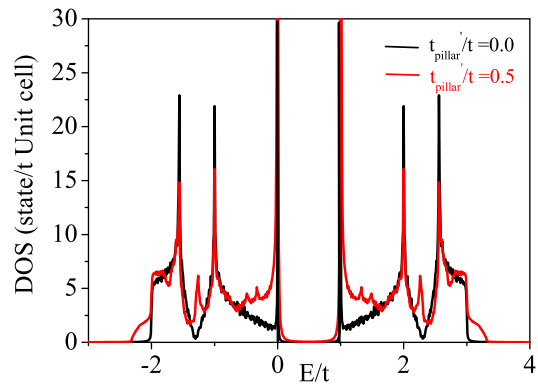}}
\caption{DOS for the structure of Fig. 2(c). 
Black and red lines represent the cases of $t^{\prime}_{pillar}/t=0.0$ and $t^{\prime}_{pillar}/t=0.5$ 
where $t^{\prime}$ is the transfer integral 
between the d$_{yz}$ (or d$_{zx}$) orbital of V and the LUMO of TCNE of the pillar. 
}
\label{Fig:8}
\end{center}
\end{figure}  

In compounds of $1.5 < x < 2$, there inevitably exists rather strong randomness 
due to the defect of V atoms.
This randomness is expected to cause the Anderson localization of the quasiparticles
at the Fermi level located in quasi-flat band~\cite{Anderson}.
Then, the T-dependence of the resistivity is expected to follow that of the VRH mechanism 
as observed in experiment~\cite{Prigodin}.

In the case of $\Delta_d/t <0$, 
the Fermi level of V(TCNE)$_2$ is located on the level of the $d_{xy}$ orbital without dispersion, 
because the $d_{xy}$ orbital is a localized one. 
In the actual system, however, the $d_{xy}$ orbital forms a narrow band, 
because it weakly hybridizes with p$_{\sigma}$ orbitals of N.
Therefore, also in this case, Kanamori mechanism arising from hidden flat bands can be considered as the cause of ferromagnetism of V(TCNE)$_x$. 

\section{Conclusion} 
In conclusion, we discussed the origin of the room temperature ferromagnetism of V(TCNE)$_x$ (1.5 $<$ $x$ $<$ 2) on the basis of the truncated model of the crystal structure newly proposed.
We estimated the exchange interaction coupling between d-electrons of V atoms by cluster model calculations. 
In parallel, the effect of itinerancy of electrons on the magnetism 
was discussed on the band picture. 
It was found that a sharp peak of the density of states appears around Fermi level 
due to the existence of the flat bands resulting from the two dimensional plane of V(TCNE)$_x$ (x=1.5 and 2), which gives rise to the ferromagnetism.

Hence the ferromagnetism of V(TCNE)$_x$ (1.5 $<$x$<$ 2) is theoretically understood both in the localized cluster model and the itinerant band picture consistently, which will indicate the fact that this ferromagnetism is very robust.
At the same time, the semiconduting transport properties (variable range hopping) in the wide range of $x$ is proposed to be due to the exsistence of that band (with the large density of state) at the fermi energy affected by the randomness.

\section*{Acknowledgements}
We are grateful to M. Tokumoto for directing our attention to
ref.\ \citen{Valade}.
One of the authors (H.M) is grateful to S. Nishiyama and Y. Yoshioka 
for their technical assistance.
This work is supported in part by a Grant-in-Aid for Specially Promoted Research (2001004)
 from The Ministry of Education, Culture, Sports, Science and Technology.
H. M. is supported by the Global COE program (G10) from The Japan Society for the Promotion of Science.

\end{document}